\documentclass[a4paper,11pt]{article}
\usepackage{pos}
\usepackage{wrapfig}
\usepackage{lineno}
\title{The Faint Particle Trigger for the IceCube Neutrino Observatory}

\author*[a]{Timo Stürwald}
\onbehalf{for the IceCube collaboration}

\affiliation[a]{Department of Physics, Faculty of Mathematics and Natural Sciences, University of Wuppertal,\\
  Gaußstraße 20, 42119 Wuppertal, Germany}

\emailAdd{stuerwald@uni-wuppertal.de}

\abstract{The Faint Particle Trigger (FPT) was developed for the IceCube Neutrino 
Observatory to enhance the detection efficiency for faint signatures, produced by 
free Fractionally Charged Particles (FCP) predicted in several Standard Model (SM) extensions. A previous IceCube analysis has shown a reduced trigger efficiency in detecting FCP with a charge of e/3 due to the $z^2$ dependence of photon production processes. The FPT addresses this shortcoming by incorporating a so far unused hit type, so called SLC hits in the trigger decision. These are single isolated hits that are not used for triggering high energy signatures in IceCube. The FPT employs a sliding time window to analyze hits, utilizing four cuts to 
remove noise and bright background contributions. The noise contribution is effectively
decreased to a few Hz by velocity and directional consistency of hit pairs. Furthermore
a part of the dominating atmospheric muon rate is reduced, by requiring a minimum 
fraction of SLC hits in the trigger window. The FPT significantly improves the trigger efficiency by a  factor of 1.55, compared to standard triggers, while increasing the event rate in IceCube by a factor 1.004. The trigger algorithm was tested and successfully deployed at South Pole in November 2023.}

\FullConference{42nd International Conference on High Energy Physics (ICHEP2024)\\
18-24 July 2024\\
Prague, Czech Republic\\}

\begin{document}
\newpage
\maketitle
\section{IceCube}
\begin{wrapfigure}{r}{0.55\textwidth} 
    \includegraphics[width=\linewidth]{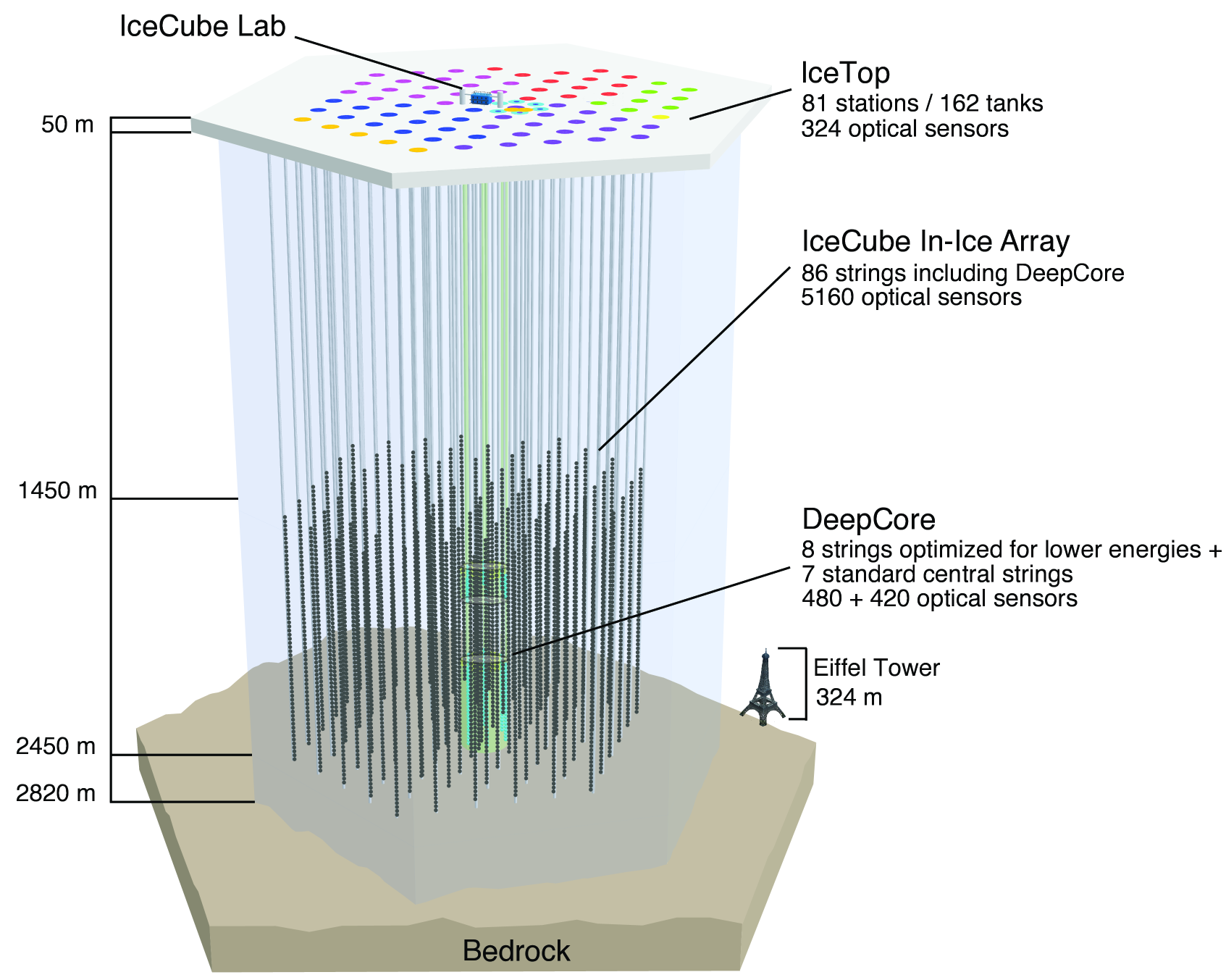} 
    \caption{The IceCube detector showing 86 strings including the highlighted DeepCore volume at the bottom.} 
    \vspace{-10pt} 
    \label{detector} 
\end{wrapfigure}

The IceCube detector is located at the geographic South Pole \cite{IceCube_online}.
It consists of 5160 Digital Optical Modules (DOMs), located 
on 86 strings, with a PMT as the main component to detect photons (\autoref{detector}). The DOMs are located at a depth between 1.45 and 2.45 km resulting in 1 $\text{km}^3$ of 
instrumented ice. DeepCore is the low energy extension of IceCube consisting of a 
subset of 15 strings, out of which 8 are DeepCore strings housing PMTs with a higher  
quantum efficiency with a reduced DOM spacing \cite{IceCube_DeepCore}. The energy threshold for the denser instrumented DeepCore volume is a few GeV while for 
IceCube it is about 100 GeV. 

All amplified PMT waveforms crossing a specific discriminator threshold result in a hit, which also stores the  position and time information. All hits are marked as SLC (Soft Local Coincidence) or HLC (Hard Local Coincidence) hit type. HLC hits are correlated hit pairs in the time window of $\pm $ 1 µs of  neighboring and next to neighboring DOMs on the same  string. SLC hits are isolated single hits. The IceCube standard triggers base their decisions on HLC hits. Each day, approximately\newline 1 TB of data are triggered, with a median trigger rate of 2.7 kHz. Around 100 GB of filtered data are transmitted via satellite per day \cite{IceCube_online}.
\section{Motivation}
 Free Fractionally Charged Particles (FCPs) appear in several extensions of the SM  with well-motivated charges being e/3, e/2 and 2e/3 \cite{frampton_fractionally_1982,barr_fractional_1983,Dong_1983}. 
They could have been produced in the early universe, in violent 
astrophysical events or in cosmic ray interactions in the upper
atmosphere and leave faint signatures as they cross the IceCube detector \cite{perl_searches_2009}. 

A previous IceCube analysis searched for FCP and for the faintest signatures with a charge of e/3 the trigger efficiency was significantly lower than for higher charges 
\cite{van_driessche_search_2019}.
This decrease in the trigger efficiency arises from the $z^2$ dependence 
affecting the production of Cherenkov photons, ionization and secondary 
processes, where $z$ is the charge. The reduced photon emission results in the 
predominance of produced SLC hits in the detector, as the probability to produce a 
correlated hit pair decreases. The Faint Particle Trigger (FPT) addresses this shortcoming by incorporating SLC hits into the trigger decision, thereby improving the overall trigger efficiency.

More recently also the search for millicharged particles at neutrino telescopes was
suggested \cite{arguelles_millicharged_2021}. These are motivated by dark matter 
models and they could be produced by various mechanisms in the upper atmosphere 
\cite{Wu2024}. The mass scale would be in the range from MeV to several GeV and IceCube could potentially probe parts of the parameter space \cite{arguelles_millicharged_2021}.
Future analyses using leptons in the GeV range and analyses of 
particles with an anomalous charge will use the data of the FPT.
\section{The Faint Particle Trigger}
A sliding time window with length of 2500 ns analyzes all DeepCore HLC and SLC hits.
Within each time window four variables are calculated and cuts are applied. 
The first three mainly aim to remove the detector noise contribution and the last cut
was constructed to keep events with a high SLC fraction, effectively reducing the 
rate of triggered atmospheric muons.
The figures in this section show distributions of cut variables and trigger efficiencies for simulations of FCP with a charge of e/3 and a mass of 1 TeV. Only high-quality events that produce at least 10 signal hits in DeepCore were selected. Additionally, simulated detector noise is shown, which consists of two main components: thermal noise and an uncorrelated component from Potassium-40 decays. Potassium-40 is present in low abundance within the spherical glass pressure vessels of the DOMs. Fixed Rate Trigger (FRT) data are also shown. The FRT reads out the entire detector for 10 ms every 300 s. These unbiased detector data were used during the development of the FPT to estimate the trigger rate and to inspect the cut variable distributions for data.

\subsection{Cut 1: Number of hits}
\begin{wrapfigure}{r}{0.57\textwidth} 
    \vspace{-40pt}
    \centering 
    \includegraphics[width=\linewidth]{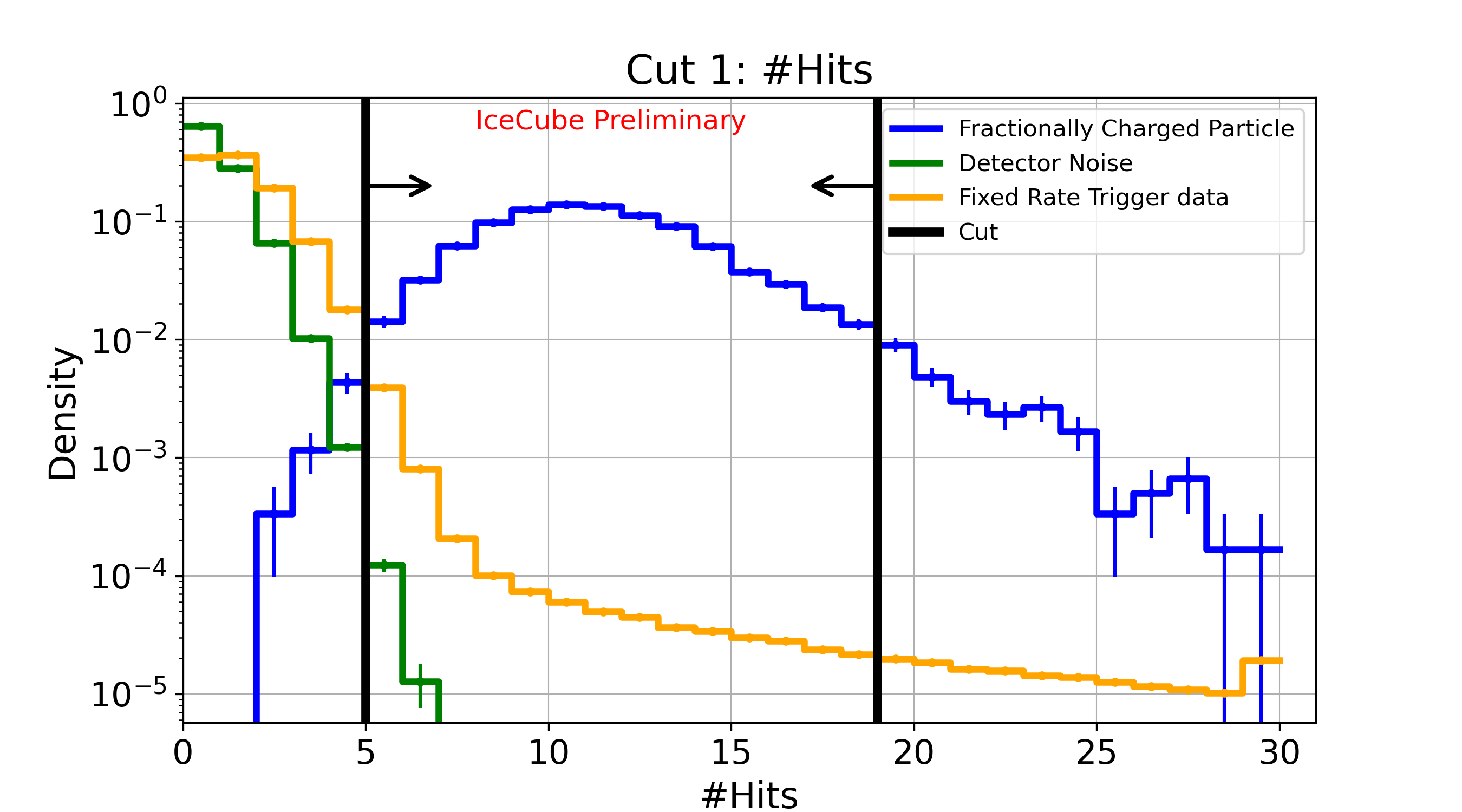}
    \caption{The number of hits in the time window. The black arrows indicate the cut values.} \label{Launches} 
    \vspace{-17pt}
\end{wrapfigure}
The first variable is the number of hits in the time window with 
corresponding distributions shown in \autoref{Launches}. One can see 
that detector noise clusters at lower values compared to the FCP.
To filter out these sparsely populated time windows, as well as those with 
an excessive number of hits that do not match the dim signature of 
interest, a threshold is set. The number of hits is required to fall 
within $5\leq \#Hits \leq 19$.
\subsection{Cut 2: Number of Doubles}
\begin{wrapfigure}{r}{0.57\textwidth} 
   \vspace{-39.5pt}
    \centering 
    \includegraphics[width=\linewidth]{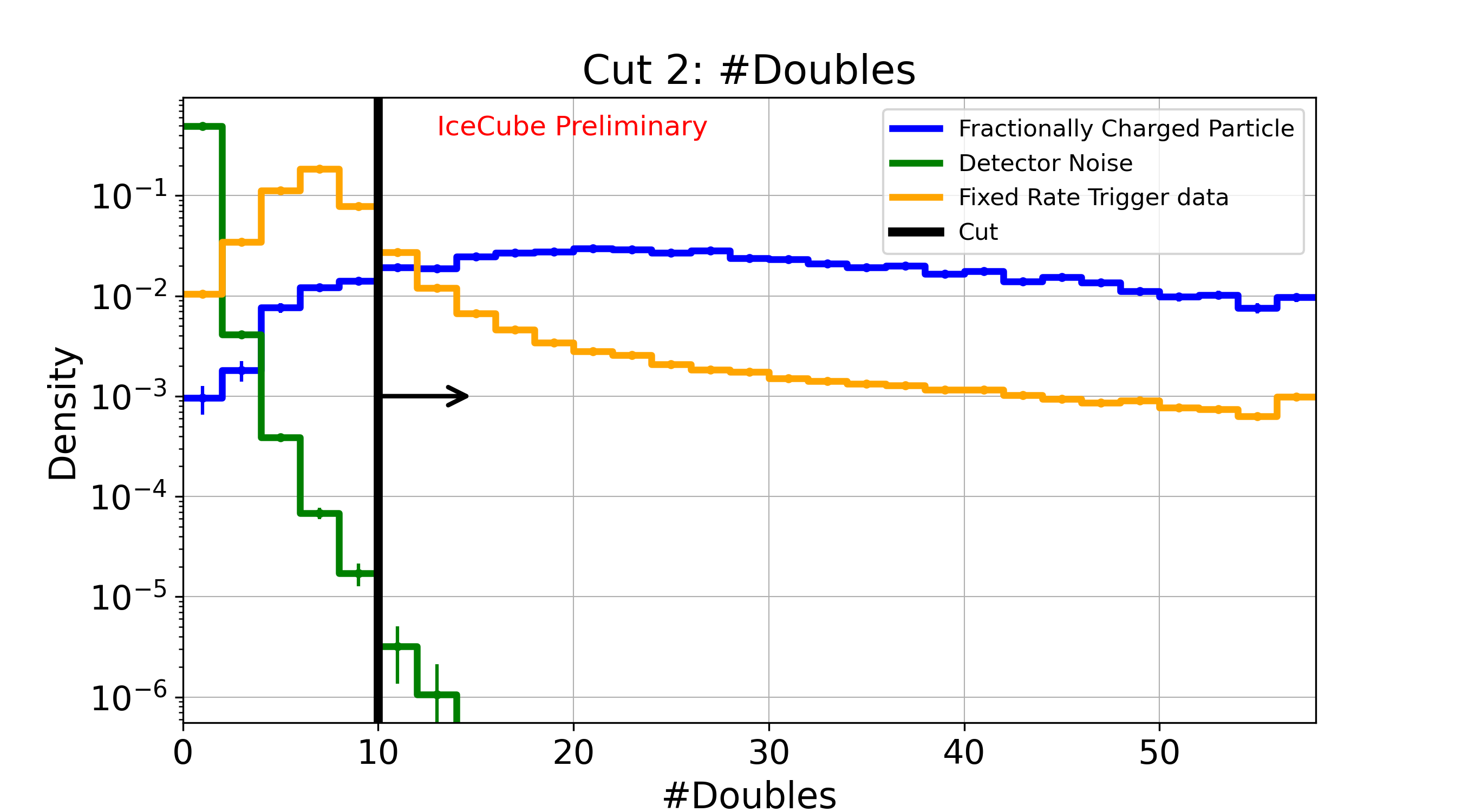}
    \caption{Number of Doubles in the time window. The black arrow indicates the cut value.} \label{Doubles} 
    \vspace{-10pt}
\end{wrapfigure}
To further reduce the rate of time windows filled with detector noise hits,
the velocity consistency of the hits in the time window is analyzed. For noise hits
no velocity consistency is expected. For this cut all hit pair combinations in the
time window are formed. Then the corresponding velocity of the hit pair
is calculated. If the velocity falls within the range $0.03 c\leq v_\text{hitpair} 
\leq 1.03 c$, the hit pair is classified as a Double. The number of Doubles (\#Doubles) per time window is counted and its corresponding distributions are shown in \autoref{Doubles}. 
The threshold is set to $\#Doubles \geq 10$ in order to reduce the detector noise contribution by more than an order of magnitude.

\subsection{Cut 3: Directional consistency of Doubles}
To further reduce the detector noise contribution the directional 
clustering of Doubles  is quantified. For detector noise no preferred 
direction is expected. For each Double in
\begin{figure}[h]
    \centering
    \begin{minipage}{0.54\textwidth}
        \centering
        \includegraphics[width=\linewidth]{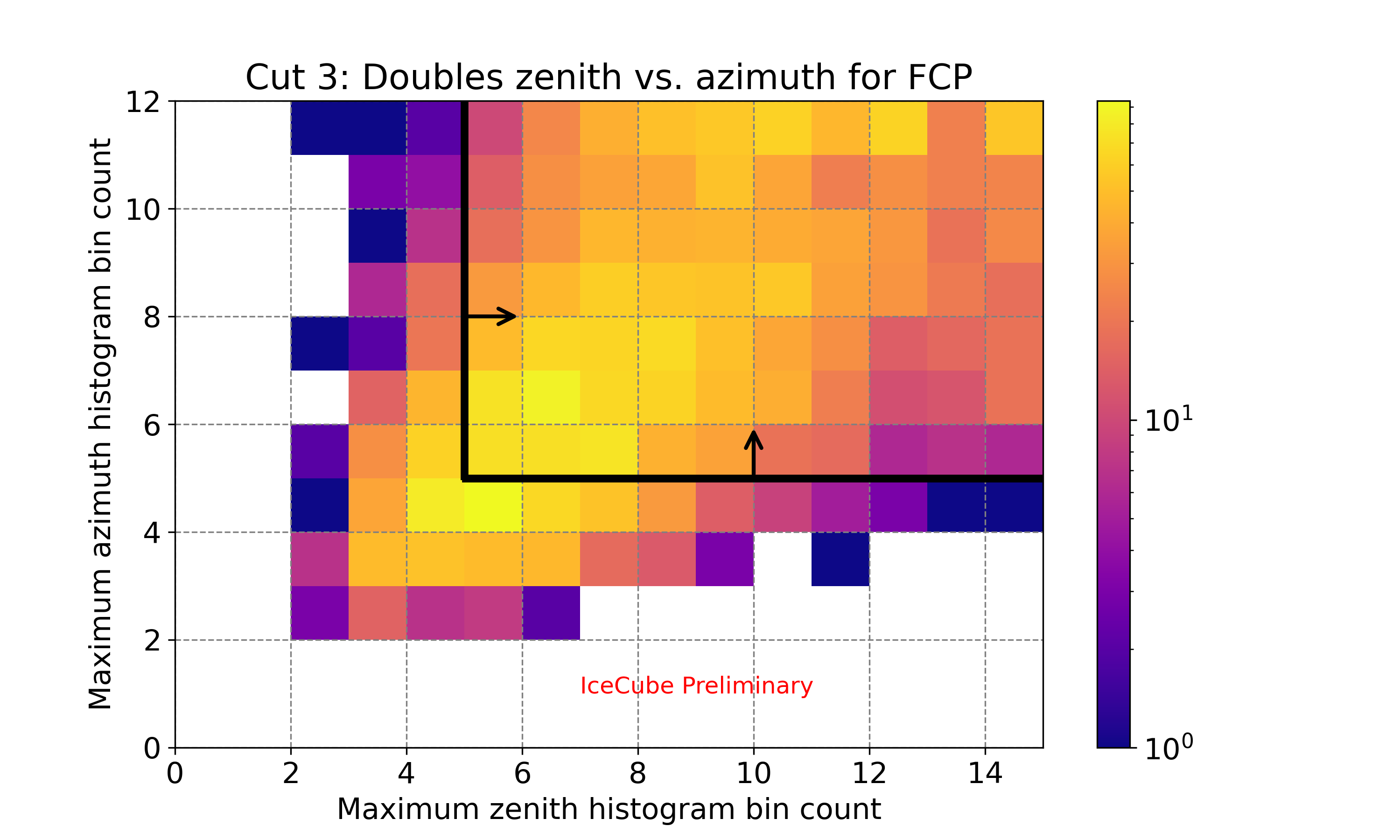}
        
    \end{minipage}
    \hspace{-40pt}
    \begin{minipage}{0.54\textwidth}
        \centering
        \includegraphics[width=\linewidth]{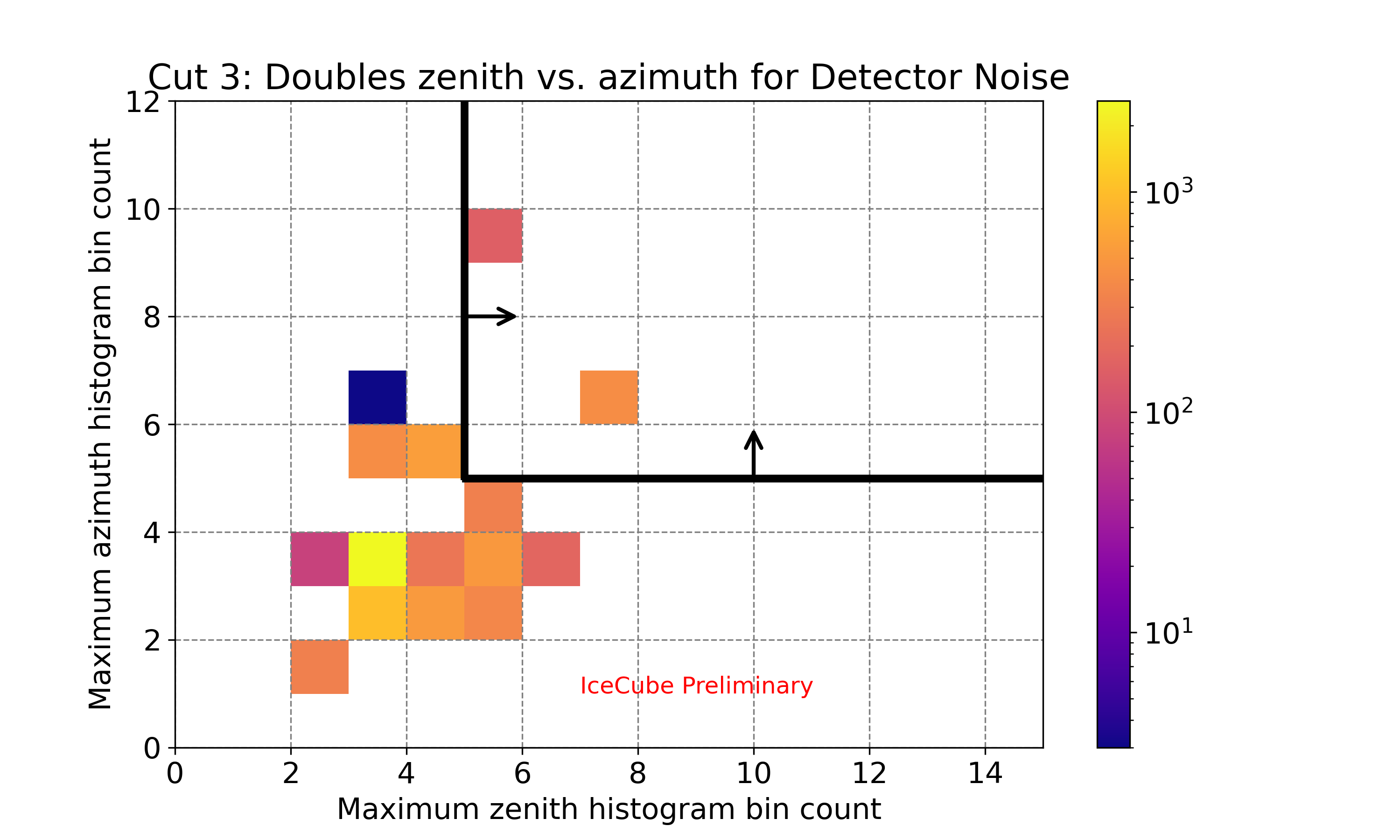}

    \end{minipage}
    \caption{The maximum zenith and azimuth histogram bin counts for FCP (left) and Detector Noise (right). The black arrows indicate the cut values.}
        \label{Doubles_direction_fcp}
\end{figure}
the time window the zenith and azimuth angle are calculated. These are binned in 20°
bins and the value of the  bin with the maximum count is compared to a threshold as 
shown in \autoref{Doubles_direction_fcp}. The detector noise values cluster at lower 
values as expected. The cut is adjusted, so that the detector noise contribution is 
reduced to a level of few Hz.

\subsection{Cut 4: SLC fraction}
\begin{wrapfigure}{r}{0.63\textwidth} 
    \vspace{-35pt}
    \centering 
    \includegraphics[width=\linewidth]{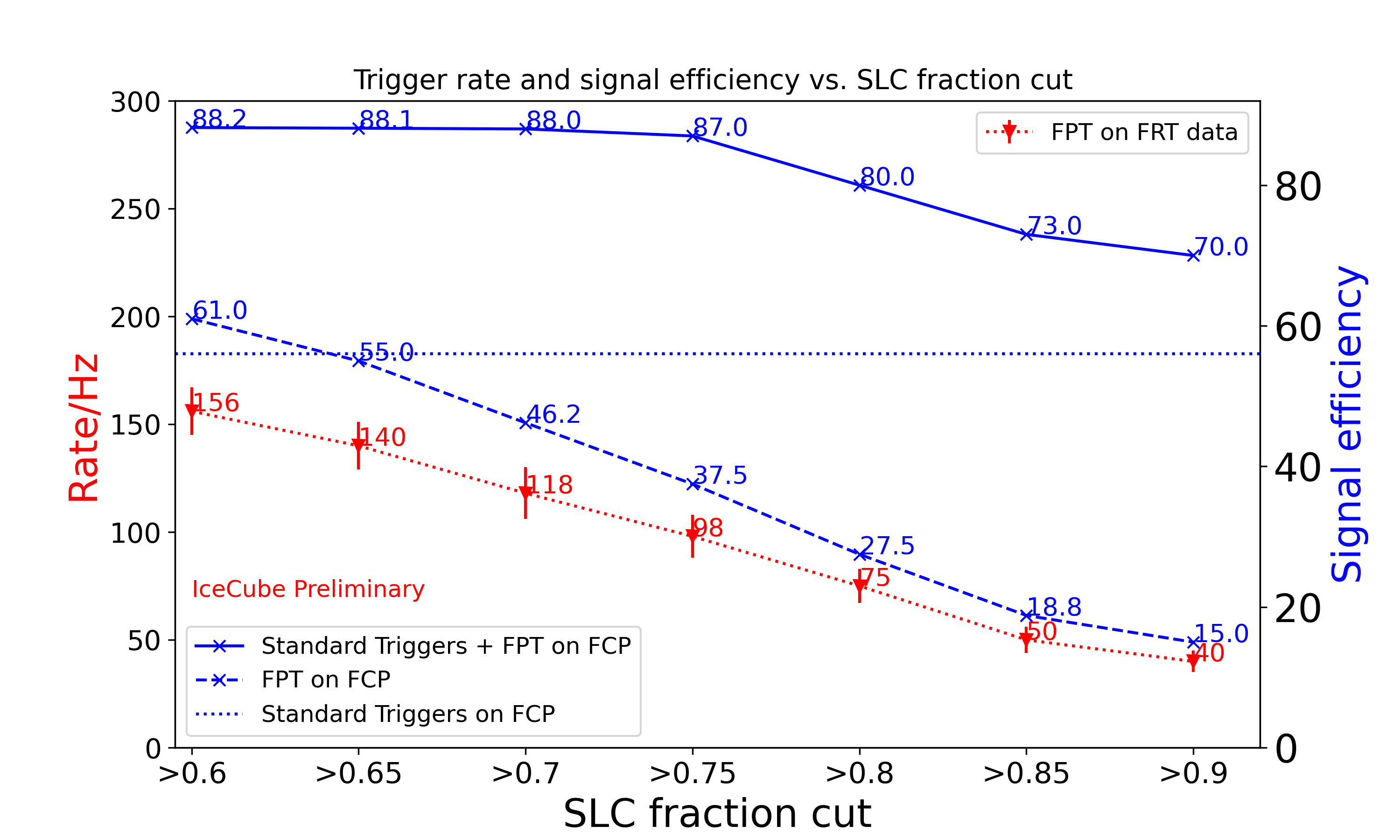}
    \caption{Estimated FPT trigger rate (dotted red). Standard trigger efficiency on FCP (dotted blue) as a reference. FPT trigger efficiency on FCP (dashed blue). Combined trigger efficiency of standard triggers and the FPT on FCP (solid blue)}
    \label{slc_cut} 
    \vspace{-10pt}
\end{wrapfigure}
As FCP dominantly produce SLC hits the last cut aims to 
remove events that predominantly produce HLC hits. For that purpose the
ratio (SLC fraction) of the number of SLC hits over all hits in the time window is 
calculated. The cut is set to $\text{SLC fraction} > 0.75$. \autoref{slc_cut} shows 
the variation of the SLC fraction cut. Following the dashed blue line one can see 
that the signal efficiency on the FCP simulation decreases rapidly when 
\begin{figure}[h]
    \centering
	\includegraphics[scale=0.45]{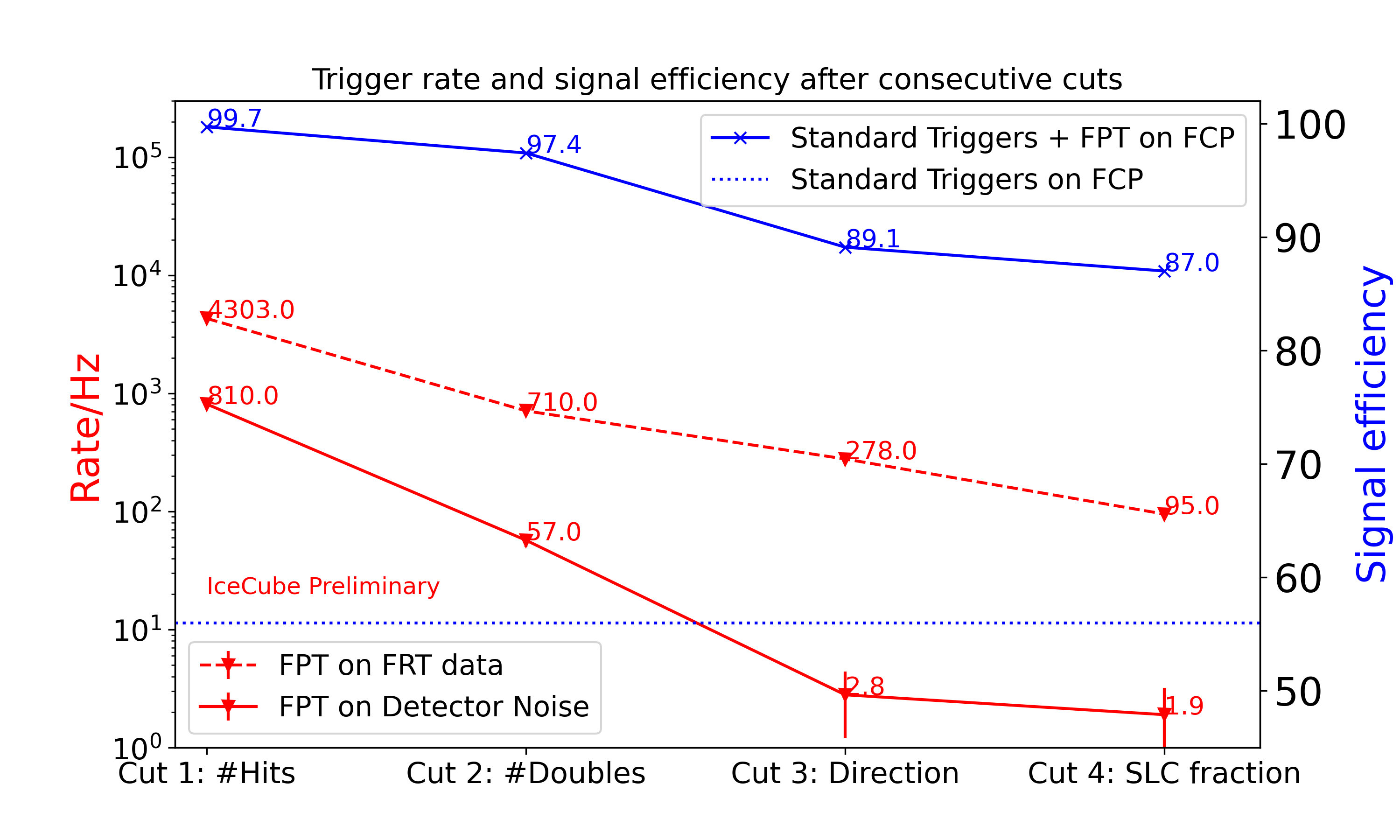}
	\caption{Summary of Cuts 1–4, applied consecutively. For each cut, the corresponding rates and signal efficiencies are shown. The blue lines show the signal efficiency from applying the standard triggers (dashed) and the standard triggers plus the FPT (solid) on FCP simulations. The red lines show the estimated FPT rates, derived from applying the FPT to detector noise simulation (dashed) and FRT data (solid).}
	\label{cut_summary}
\end{figure}
\noindent
applying the FPT to them. Nevertheless, the combined signal efficiency (solid blue)
is stable up to a value of 0.75. Thus, FCP events in the SLC fraction region 0.6 - 0.75
are mostly already covered by the standard triggers. This shows that additionally 
triggered events by the FPT typically have a high SLC fraction above 0.75. Setting 
the threshold to >0.75 reduces the FPT rate to approximately 100 Hz (dotted red).
\subsection{Summary of cuts and results}
\begin{wraptable}{r}{0.35\textwidth}  
    \vspace{-12pt}
    \begin{tabular}{ |p{1.4cm}|p{2.9cm}|  }
     \hline
     Flavor \& energy & Relative trigger\newline efficiency increase \\
     \hline
    $\nu_{e,A}$ & 1.11 $\pm\,$ 0.02 \\
    $\nu_{e,B}$ & 1.18 $\pm\,$ 0.02 \\
    $\nu_{e,C}$ & 1.10 $\pm\,$ 0.01 \\
    $\nu_{\mu,A}$ & 1.11 $\pm\,$ 0.02 \\
    $\nu_{\mu,B}$ & 1.10 $\pm\,$ 0.01 \\
    $\nu_{\tau,A}$ & 1.14 $\pm\,$ 0.02 \\
    $\nu_{\tau,B}$ & 1.15 $\pm\,$ 0.01 \\
     \hline
    \end{tabular}
    \caption{Relative improvements above 1.03 for neutrino simulation in DeepCore resulting from including the FPT. The energy intervals are: A(1--4~GeV),  B(4--12~GeV) and \newline C(12--100~GeV).}
    \label{table1}
    \vspace{-10pt}  
\end{wraptable}
The impact of each cut on the signal efficiency, detector noise rate, and
the estimated trigger rate is summarized in \autoref{cut_summary}. 
After applying the fourth cut, 87\%  of the signal events are 
retained (solid blue), representing a relative improvement of 1.55 compared to 
the standard triggered events. It can be observed that the contribution from detector noise is reduced to approximately a few Hz after the third cut (solid red). 
The fourth cut does not significantly further reduce the noise contribution, because detector noise predominantly produces SLC hits. On the other hand, the estimated FPT rate (dashed red) is reduced by an additional factor of three by the fourth cut. This suggests that the fourth cut is particularly effective in removing muons, which typically produce events with a lower SLC fraction compared to FCP. The trigger was applied to existing data sets of electron, muon and tau neutrino simulation in DeepCore. The relative improvements are shown in \autoref{table1}. The largest improvement of 1.18 can be found for electron neutrinos between 4-12 GeV. 

\section{Test and deployment}

The trigger algorithm was tested at the South Pole Test System, IceCube’s test DAQ 
facility. This setup allows for the testing of newly developed software. The trigger
algorithm was tested on a IceCube standard run, which has a typical duration of 8 
hours. The FPT consumes approximately 1\% of the CPU resources and runs stable. An 
event in IceCube can consist of multiple triggers that fire close 
in time. The event rate for the test run increases from $2784.0\pm\, 0.3$ Hz to  
$2793.9\pm\, 0.3$ Hz by adding the FPT. The additional 10 Hz of events are only 
triggered by the FPT and correspond to approximately 3.4 GB of additional data per day.
\begin{wrapfigure}{r}{0.71\textwidth}
	\includegraphics[width=\linewidth]{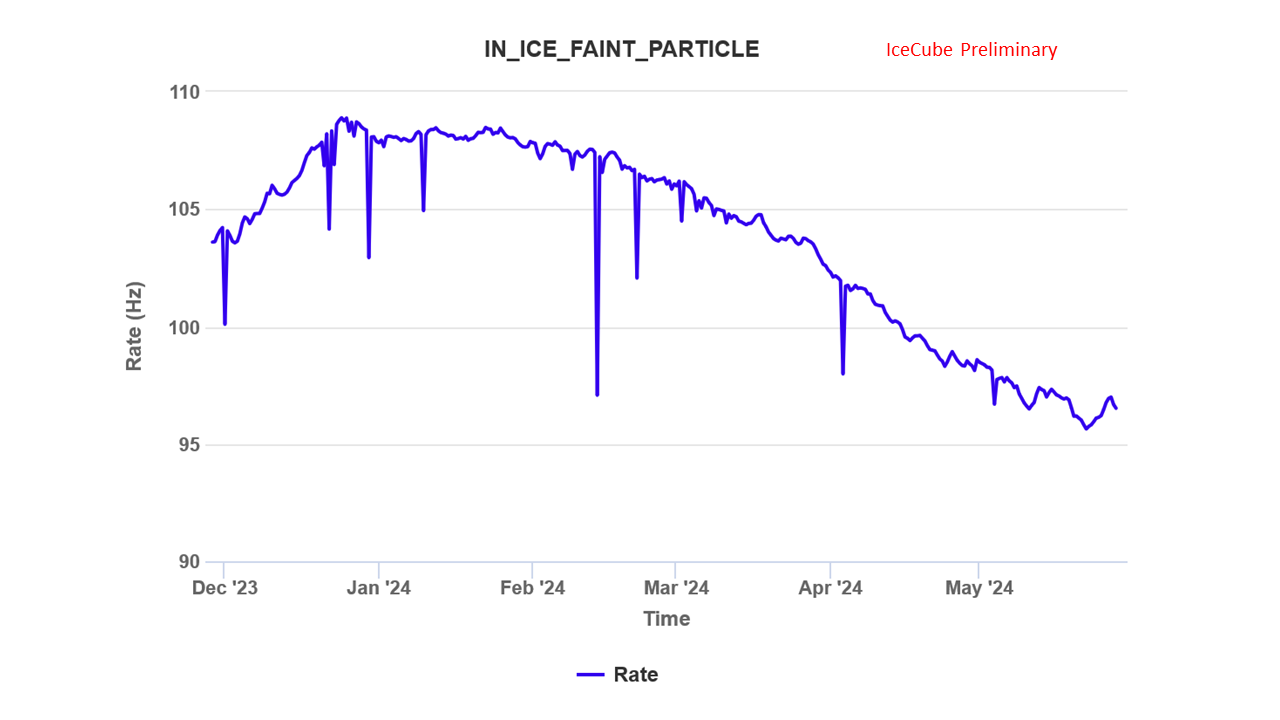} 
	\centering
	\caption{The FPT rate for the first six months of operation.}
	\label{fpt_rate}
    \vspace{-10pt}
\end{wrapfigure}
The relative event rate increase corresponds to a factor of 1.004. The trigger was
successfully deployed at the South Pole on November 28 2023. The rate of the first six 
month can be seen in \autoref{fpt_rate}. The rate trend aligns with the trend of the other main triggers which is correlated to seasonal variations of the atmospheric temperature \cite{Tilav2009}.
\section{Conclusion}
The FPT utilizes the information of the so far unused SLC hits for the trigger decision, enhancing sensitivity to faint signatures in IceCube up to a factor of 1.55. The successful deployment of the FPT at the South Pole results in approximately 10 Hz of additional events suitable for analyses involving leptons with energies up to 100 GeV and beyond the SM searches.


\begin{thebibliography}{99}
{\fontsize{9}{1}\selectfont
\bibitem{frampton_fractionally_1982}
P. H. Frampton and T. W. Kephart, \textit{Fractionally Charged Particles as Evidence for Supersymmetry}, \href{https://journals.aps.org/prl/abstract/10.1103/PhysRevLett.49.1310}{Phys. Rev. Lett. \textbf{49}, 1310--1313 (1982)}.

\bibitem{barr_fractional_1983}
S. M. Barr, D. B. Reiss, and A. Zee, \textit{Fractional Charges, Monopoles, and Peculiar Photons}, \href{https://journals.aps.org/prl/abstract/10.1103/PhysRevLett.50.317}{Phys. Rev. Lett. \textbf{50}, 317--320 (1983)}.

\bibitem{Dong_1983}
F.-X. Dong, T.-S. Tu, P.-Y. Xue, and X.-J. Zhou, \textit{Fractional charges,
monopoles and peculiar photons in SO(18) GUT models}, 
\href{https://doi.org/10.1016/0370-2693(83)90129-6}{Phys. Lett. B \textbf{129}, 405--410 (1983)}.

\bibitem{van_driessche_search_2019}
W. Van Driessche, \emph{Search for particles with anomalous charge in the IceCube detector}, PhD dissertation, Ghent University, 2019, \url{https://biblio.ugent.be/publication/8683108}.

\bibitem{arguelles_millicharged_2021}
C. A. Argüelles, K. J. Kelly, and V. M. Muñoz, \textit{Millicharged Particles from the Heavens: Single- and Multiple-Scattering Signatures}, \href{https://doi.org/10.1007/JHEP11(2021)099}{JHEP, \textbf{2021}, 99 (2021)} \href{https://arxiv.org/abs/2104.13924}{[2104.13924]}.
\bibitem{IceCube_online}
IceCube Collaboration,
\textit{The IceCube Neutrino Observatory: Instrumentation and Online Systems}, 
\href{https://doi.org/10.1088/1748-0221/12/03/P03012}{JINST \textbf{12}, P03012 (2017)} 
\href{https://arxiv.org/abs/1612.05093}{[1612.05093]}.

\bibitem{IceCube_DeepCore}
IceCube Collaboration,
\textit{The Design and Performance of IceCube DeepCore},
\href{https://doi.org/10.1016/j.astropartphys.2012.01.004}{Astropart. Phys. \textbf{35}, 615 (2012)},
 \href{https://arxiv.org/abs/1109.6096}{[1109.6096]}.

\bibitem{perl_searches_2009}
M. L. Perl, E. R. Lee, and D. Loomba, \textit{Searches for Fractionally Charged Particles}, \href{https://doi.org/10.1146/annurev-nucl-121908-122035}{Annual Review of Nuclear and Particle Science \textbf{59}, 47-65 (2009)}
\bibitem{Tilav2009}
 IceCube Collaboration,
\textit{Atmospheric Variations as Observed by IceCube},
July 2009, \href{https://doi.org/10.48550/arXiv.1001.0776}{1001.0776}.

\bibitem{Wu2024}
H. Wu, E. Hardy, and N. Song
\textit{Searching for heavy millicharged particles from the atmosphere},
  June 2024, \href{https://arxiv.org/abs/2406.01668}{2406.01668}.

}
\end{thebibliography}
\end{document}